\documentclass[aps,pra,superscriptaddress,twocolumn]{revtex4-1}
\usepackage{blindtext}
\usepackage{bm}
\usepackage{amsmath,amssymb,amsfonts,latexsym,fancyhdr,graphicx,times,txfonts}
\usepackage{simplewick}
\usepackage[english]{babel}
\usepackage{graphicx}
\usepackage{subfigure}
\usepackage{xcolor}
\usepackage{color}
\usepackage{verbatim}
\usepackage[pdftex]{hyperref}
\usepackage{enumitem}
\usepackage{dsfont}

\newcommand{\No}[1]{}

\newcommand{\ketbra}[2]{\vert#1\rangle\langle#2\vert}
\newcommand{\braketM}[3]{\langle#1\vert#2\vert#3\rangle}

\begin{document}

\title{Emergence of anomalous dynamics \\from the underlying singular continuous spectrum in interacting many-body systems}

\author{J. Settino}
\affiliation{Dipartimento di Fisica, Università della Calabria, 87036
Arcavacata di Rende, Cosenza, Italy}
\affiliation{INFN, Gruppo collegato di Cosenza, 87036 Arcavacata di Rende,
Cosenza, Italy}
\author{N. W. Talarico}
\affiliation{QTF Centre of Excellence, Turku Centre for Quantum Physics,
Department of Physics and Astronomy, University of Turku, 20014 Turku, Finland}
\author{F. Cosco}
\affiliation{QTF Centre of Excellence, Turku Centre for Quantum Physics,
Department of Physics and Astronomy, University of Turku, 20014 Turku, Finland}
\author{F. Plastina}
 \affiliation{Dipartimento di Fisica, Università della Calabria, 87036
 Arcavacata di Rende, Cosenza, Italy}
 \affiliation{INFN, Gruppo collegato di Cosenza, 87036 Arcavacata di Rende,
 Cosenza, Italy}
\author{S. Maniscalco}
\affiliation{QTF Centre of Excellence, Turku Centre for Quantum Physics,
Department of Physics and Astronomy, University of Turku, 20014 Turku, Finland}
\affiliation{ QTF Centre of Excellence, Department of Applied Physics, Aalto University, FI-00076 Aalto, Finland}
\author{N. Lo Gullo}
\affiliation{QTF Centre of Excellence, Turku Centre for Quantum Physics,
Department of Physics and Astronomy, University of Turku, 20014 Turku, Finland}

\begin{abstract}
We investigate the dynamical properties of an interacting many-body system with a non-trivial energy potential landscape that may induce a singular continuous single-particle energy spectrum. Focusing on the Aubry-Andr\'e model, whose anomalous transport properties in presence of interaction has recently been demonstrated experimentally in an ultracold gas setup, we discuss the anomalous slowing down of the dynamics it exhibits and show that it emerges from the singular-continuous nature of the single-particle excitation spectrum.
Our study demonstrates that singular-continuous spectra can be found in interacting systems, unlike previously conjectured by treating the interactions in the mean-field approximation.
This, in turns, also highlights the importance of the many-body correlations in giving rise to anomalous dynamics, which, in many-body systems, can result from a non-trivial interplay between geometry and interactions.
\end{abstract}
\selectlanguage{english}
\maketitle

\section{Introduction}
The discovery of quasicrystals in 1982~\cite{Shechtman1984} and  of protocols to produce large and stable samples~\cite{Tsai1987}
has triggered theoretical investigations to clarify the origin of their unusual physical properties such as increase of resistivity with both decreasing temperature and/or increasing the sample purity \cite{dubois}.
It was soon realized that this behavior is strictly linked to the singular continuous (SC) nature of the single-particle energy spectrum (SPES), with the accompanying critical eigenfunctions~\cite{note1}, whose scaling properties account for anomalous transport and diffusion~\cite{Roche1997}, and partially explain the peculiar behavior of these materials.
Before the discovery of quasicristalline structures, the SC spectrum~\cite{note2} was thought to be solely a mathematical concept with no physical counterpart~\cite{Steinhardt2013}.
The SC part, in fact, is not easily accessible and often its presence is inferred
after removing the absolutely continuous (AC) and pure-point (PP) parts from the whole spectrum, provided a set of non-zero measure is left over.

The role of SC spectra in the dynamics of non-interacting systems has been investigated in Ref.~\cite{Zhong1995} and its link to anomalous propagation of correlations and to the spreading of an initially localized wave-packet has been investigated in Refs.~\cite{Ketzmerick1992,LoGullo2017}. A particularly interesting, exemplary physical model where the nature of the spectrum plays a crucial role is the Aubry-Andr\'e model (AAM), which describes particle hopping in a one-dimensional quasi-periodic lattice.
It displays a metal-to-insulator transition~\cite{Jitomirskaya1999,Jitomirskaya2009}, with the spectrum being AC and PP in the metal and insulating phases, respectively, while it is purely SC at the transition point. The model has been realized with ultra-cold atoms loaded in a bichromatic optical lattice \cite{modugno,Ulrich2015,Ulrich2017}.
Due to the interplay of quasi-periodicity and inter-particle interaction, a non-trivial phase diagram arises~\cite{Roscilde2008,Deng2008,Roux2008}, together with the appearance of a mobility edge ~\cite{Naldesi2016,Settino2017,Ancilotto2018}, a many-body-localized phase~\cite{Ulrich2015,Ulrich2017,Prelovsek2016,Mace2018,dassarma19}, and instabilities~\cite{Znidaric2018}. The problem of how interactions modify the properties of SC spectra has been addressed in the seminal work~\cite{Kohomoto1992}, with the conclusion that they would destroy SC SPES. However, in Ref.~\cite{Kohomoto1992}, interactions have been treated in the mean-field approximation, and correlation effects were not included.
The same behavior has been found in Ref.~\cite{LoGullo2015}, where boson-boson interactions have been treated within the Bogoliubov approximation which is an effectively non-interacting theory.

Inspired by recent experiments~\cite{Ulrich2015,Ulrich2017}, we provide an explanation of the observed dynamical slowing down of an interacting gas loaded in an incommensurate bichromatic lattice, which is based on the nature of the SPES.
We find different dynamical regimes for the system: an ergodic one for small values of the amplitude of the quasi-periodic potential modulation (called $\lambda$, below) with an AC SPES, and a localized one at large $\lambda$'s and moderately small interactions with a PP SPES.
These two extreme behaviors are separated by an intermediate region, characterized by a SC SPES, where the dynamics is still
ergodic, but on time scales much longer than the typical single-particle ones.
Our findings imply that a non-trivial competition takes place between the underlying order induced by the potential energy landscape and the many-body interaction.

\section{The model}
We consider a gas of spin-$1/2$ particles in one dimension, described
by the Fermi-Hubbard model:
\begin{equation}
 \hat{H}= J \sum\limits_{n,\sigma} \epsilon_{n}\hat{c}_{n,\sigma}^{\dag}\hat{c}_{n,\sigma}
 -\frac{1}{2}\left(\hat{c}_{n,+1,\sigma}^{\dag}\hat{c}_{n,\sigma}+\text{h.c.}\right)
 +U\;\hat{n}_{n,\uparrow}\hat{n}_{n,\downarrow},
 \label{eq:ham}
\end{equation}
where  $J$ is the hopping strength (which we use as energy unit), $\epsilon_{n}$ is the onsite energy, $U$ the on-site interaction between particles with different spin in the s-wave approximation, $\hat{c}_{n,\sigma}^{\dag}(\hat{c}_{n,\sigma})$ are fermion creation (annihilation) operators at site $n$ with spin $\sigma$ and $\hat{n}_{n,\sigma}=\hat{c}_{n,\sigma}^{\dag}\hat{c}_{n,\sigma}$ the corresponding number operator. We choose to work with open boundary conditions not to enforce any artificial periodicity. The AAM is obtained by setting $\epsilon_{n}= \lambda\cos(2\pi\tau n)$~\cite{note3} with $\tau=\left(\sqrt{5}+1\right)/2$.

If not otherwise stated, we consider an initial state with two particles with opposite spin on even sites with odd sites being empty which can be considered as the ground state of an Hamiltonian with no interaction and shallower on-site potential on even sites. At time t=0 we assume that a sudden quench of the interaction and of the on-site potential realizes the Hamiltonian $\hat{H}$  which remains constant in time and governs the dynamics of the system.
It is important, for the forthcoming discussion, to mention that the dynamical behaviour of the system is essentially independent of the choice of the initial state, provided the latter is spread among most of the eigenstates in the delocalized region. This guarantees that during the time evolution the system can explore most of the spectrum. This has already been exploited in Ref.~\cite{LoGullo2017} for the case of a quantum walk in aperiodic lattices.

Information on the dynamical properties of the system are obtained from the lesser component of the single-particle Green's function: $G^{<}_{ss'}(t;t')=i \langle\hat{c}_{s'}^{\dag}(t')\hat{c}_{s}(t)\rangle_0$, where the average is over the initial state, and $s={n,\sigma}$. Its time-diagonal component is nothing but the reduced single particle density matrix of the system up to a factor $i$, whereas the off.diagonal ones give information on the correlations developed during the evolution across the system.
We resort to the non-equilibrium Green's functions technique, by solving numerically the corresponding set of Dyson equations~\cite{Talarico2019}:
\begin{eqnarray}
	G^R(t;t')&=&g^R_{ss'}(t;t)+[g^R\circ \Sigma^R\circ G^R](t;t')\\
	G^{\lessgtr}(t;t')&=&G^R(t;t_0)\cdot g^{\lessgtr}(t_0;t_0)\cdot G^A(t_0;t)\\
	&&+[G^R\circ \Sigma^{\lessgtr}\circ G^A](t;t')\nonumber
\end{eqnarray}
where $[A\circ B](t;t')= \int d\bar{t}\; A(t;\bar{t})\cdot B(\bar{t};t')$ and "$\cdot$" is the matrix multiplication.
The self-energy entering the Dyson equation is calculated in the second-Born approximation~\cite{stefleebook}:
\begin{eqnarray}
\Sigma^{\lessgtr}(t;t')&=&G_{ij}^{\lessgtr}(t;t')\sum_{kl}\textrm{v}_{ik}G_{kl}^{\lessgtr}(t;t')G_{lk}^{\gtrless}(t';t)\textrm{v}_{lj}\\
\Sigma^R(t;t')&=&\Sigma_{\text{H}}(t;t')+\Theta(t-t')\left(G^{>}(t;t')-G^{<}(t;t')\right)
\end{eqnarray}
where $\textrm{v}_{ij}=U\delta_{nn'}(1-\delta_{\sigma \sigma'})$ is the on site interaction between particles with opposite spin component and $\Sigma_H(t;t')=-i\textrm{v}\cdot G^{<}(t;t)$ is the Hartree self-energy (mean field) which is local in time and depends upon the density. The Fock and the exchange diagram of the second Born are absent because of the absence of initial correlations between the spins degree of freedom and because the Hamiltonian does not create such correlations.
Our approach closely follows Refs.~\cite{vanLeeuwen2009,vanLeeuwen2016}
and is an extension of the self-consistent approach presented in Ref.~\cite{LoGullo2016} for bosonic systems. 
The self-consistency guarantees that the macroscopic conservation laws are satisfies because the Second-Born self-energy can be derived from a Luttinger-Ward potential~\cite{stefleebook}.
In the following we will also look at the spectral function $A(k,\omega)=-\pi^{-1}\textrm{Im}\;G^R(k,\omega)$ where

\begin{equation}
	G^R(k,\omega)=\underset{T\rightarrow \infty}{\lim}\sum\limits_{nm \sigma}\frac{e^{-ik(n-m)}}{2\pi}\int\limits_{-\infty}^{\infty}d\tau \; G^R_{n\sigma m\sigma}(T+\tau/2;T-\tau/2).
\end{equation}
In our numerical calculations we will choose $T$ to be half of the total time of evolution.

\section{Geometry-induced anomalous diffusion}
\begin{figure}[t]
\includegraphics[width=0.9\linewidth]{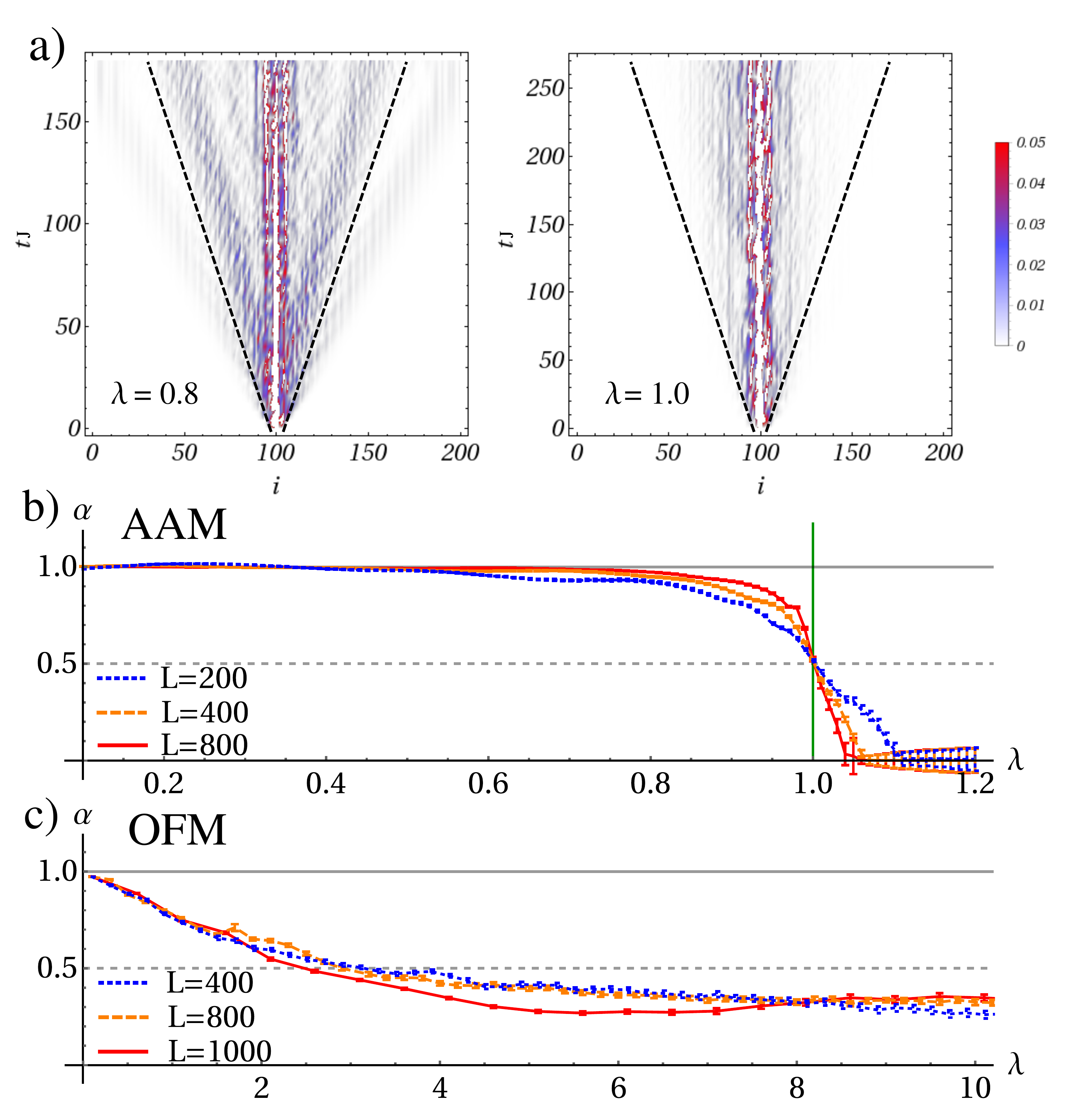}
 \caption{(Color online). {\bf a)} Spreading of the single particle correlations quantified by $|G^{<}_{i_0i}(0;t)|^2$ with $i_0=100$ for the AAM and two values of the on-site potential $\lambda$.
 Panels {\bf b)} and {\bf c)} show the exponent of the power law $\sigma(t)\propto t^\alpha$
as a function of the on-site potential strength $\lambda$. {\bf b)} refers to the AAM with
 $L=200$ (dotted blue line), $L=400$ (dashed orange line), $L=800$ (solid red line),
 while {\bf c)} refers to the OFM with $L=400$ (dotted blue line), $L=800$ (dashed orange line),$L=1000$ (solid red line).
 Solid and dashed horizontal lines highlight the values of the exponent
 $\alpha_B=1$ , $\alpha_D=0.5$ expected for ballistic and diffusive spreading, respectively.
The vertical green line at $\lambda=1$ in panel {\bf a)} signals the metal-to-insulator transition point of the AAM, in the thermodynamic limit.}
\label{fig:noninter}
\end{figure}
The spreading of correlations in a non-interacting system with an AC SPES is ballistic with a maximum velocity determined by both the energy spectrum
and the initial state (but always bounded from above by the Lieb-Robinson bound~\cite{Lieb1972}).
In the case of a PP SPES, instead, the spreading is suppressed and correlations develop only in a finite region whose size is proportional to the localization length which vanishes in the thermodynamic limit.
\begin{figure*}[t]
	\vspace{0.5cm}
	\includegraphics[width=\linewidth]{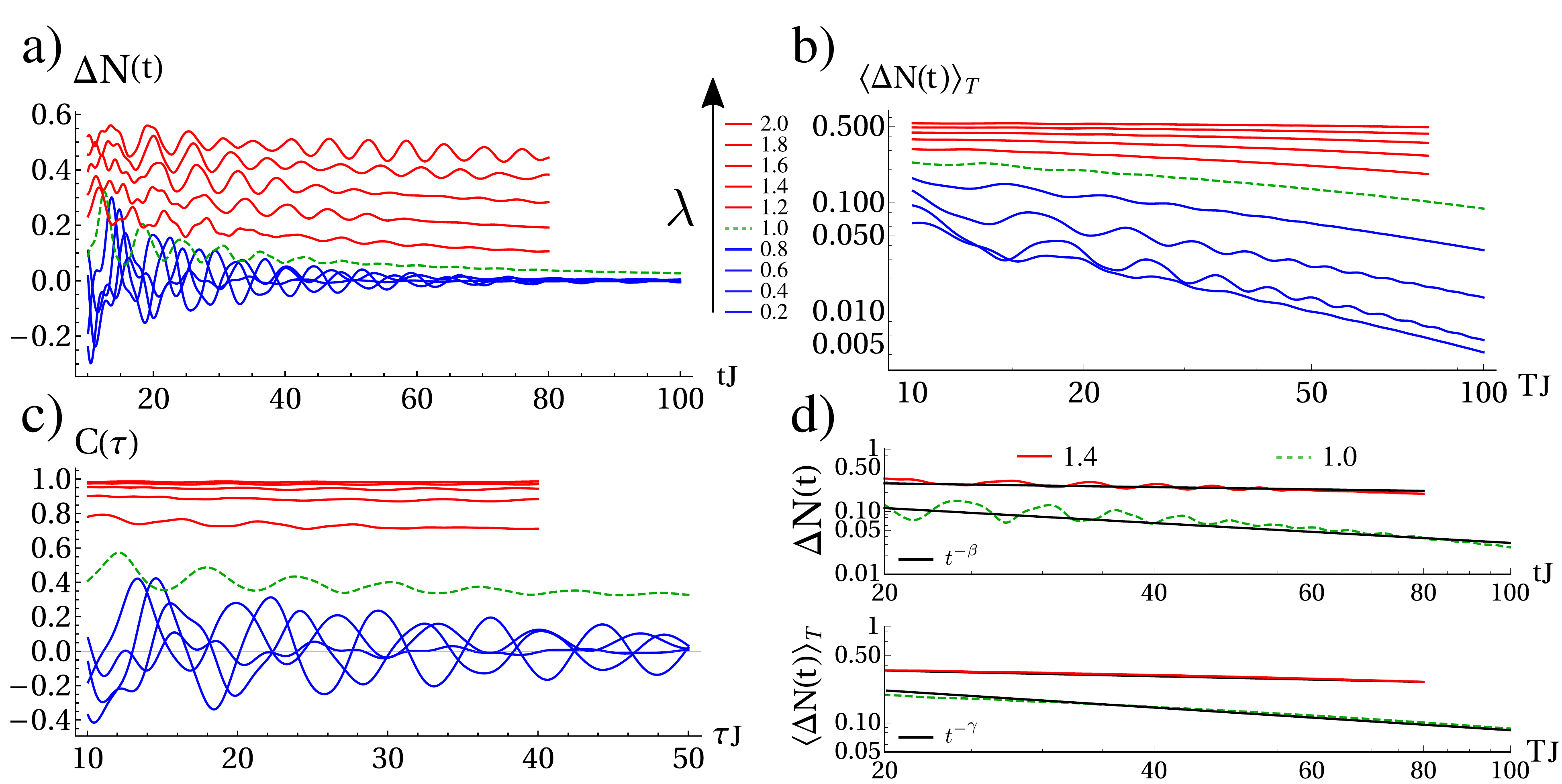}
	\caption{(Color online): {\bf a)} Particle imbalance between even and odd sites, $\Delta N(t)$, {\bf b)} Time-average $\langle \Delta N(t)\rangle_T$ (Log-Log).  {\bf c)} Autocorrelation function for a system with $L=40$ sites in the interacting AAM for $U=0.4 $ for different values of $\lambda$.
		{\bf d)} Power-law fit (solid black) for (top) $\Delta N(t)$ and (bottom) $\langle \Delta N(t)\rangle_T$ together with the data for the cases (dashed green) $\lambda=1.0$ and (solid red) $\lambda=1.4$.}
	\label{fig:slowAA}
\end{figure*}

To quantify the spreading of the correlations, we use the variance of the probability distribution defined as $P_i(t)=|G^{<}_{i_0,i}(0;t)|^2$ as in Ref.~\cite{LoGullo2016}. In Fig.~\ref{fig:noninter} panel {\bf a)} we show two examples of the $P_i(t)$ for two different values of $\lambda$, one corresponding  Due to the absence of interaction, the spin degree of freedom is irrelevant, therefore we consider spinless fermions when $U=0$.
By assuming a power-law behaviour for the variance of $P_i(t)$ at long times, i.e. $\sigma(t) \propto t^\alpha$ for $J t \gg 1$, we looked at the behaviour of the exponent $\alpha$ for different system sizes and different values of $\lambda$. The results are shown in Fig. \ref{fig:noninter} panel {\bf b)}.
In the thermodynamic limit, the expansion is ballistic ($\alpha=1$) for $\lambda<1$, whereas it is suppressed for $\lambda>1$~\cite{notelocexp}.

It is interesting to compare these features with those of the on-site Fibonacci model (OFM), showing a purely SC energy spectrum~\cite{Suto1989} induced by its quasiperiodic geometry~\cite{LoGullo2016a,LoGullo2017a}, and displaying no phase transition. The OFM is obtained by setting $\epsilon_{n}= \lambda(\lfloor (n+1)/\tau\rfloor-\lfloor n/\tau\rfloor)$ in Eq.(~\ref{eq:ham}). The results are shown in Fig.~\ref{fig:noninter} panel {\bf c)}, where we can appreciate a deviation
from ballistic spreading at any finite $\lambda$.
This behavior can be traced back to the critical nature of the eigenfunctions together with the SC nature of
the spectrum~\cite{Suto1989,Roche1997,LoGullo2017}, and it is shared by other aperiodic structures~\cite{Luck1989,Queffelec2010}.

We can draw two main conclusions from the above observations.
The AAM for $\lambda<1 (\lambda>1)$ behaves as any other non-interacting system with an AC (PP) SPES, inducing ballistic (suppression of) propagation of correlations.
At the transition point ($\lambda=1$) the AAM shares with the OFM the SC nature of the SPES,
which induces a deviation from either a simple ballistic propagation or full localization.

\section{Interplay between interaction and geometry} On-site interactions alter transport properties in a substantial way: when the single-particle eigenfunctions are extended, the spreading turns from ballistic to diffusive for moderate values of $U$~\cite{Bloch2013,LoGullo2016}; instead, in the localized case interactions help the system to acquire a non-zero diffusivity.

We have shown that anomalous diffusion arises in a non-interacting system, due to quasi-periodicity. Therefore, it is meaningful to ask how these features, induced by a non-trivial underlying geometry, are affected by the interaction. To answer this question, we look at the dynamics of a many-body interacting system described by the Hamiltonian in Eq.(~\ref{eq:ham}) both for the AAM and the OFM.

We introduce the particle imbalance, defined as
$\Delta N(t) = (N_e(t)-N_o(t))/N_{tot}$, where $N_{e/o}(t)$ is the number of particles
at the even$/$odd sites at time $t$ and $N_{tot}$ is the total number of particles in the system.
This is an experimentally accessible physical quantity~\cite{Bloch2013,Ulrich2015,Ulrich2017} and it is a good figure-of-merit for the diffusion properties of a system. In the delocalized (ergodic) phase $\Delta N(t)\rightarrow 0$ on a single-particle time scale ($\sim J^{-1}$) and all particles are redistributed among different sites. In the localized phase, $\Delta N(t)\rightarrow \bar N(\lambda,U)\ne0$ at long times $(Jt\gg 1)$. In Refs.~\cite{Ulrich2015,Ulrich2017} it has been shown that this is true away from the zero-interaction transition point. Close to $\lambda=1$, $\Delta N\rightarrow 0$ with a power-law behavior. The latter is a signature of a non-trivial interplay between the effect of interaction and geometry that we want to investigate here in more detail.

\begin{figure}[t]
	\includegraphics[width=\linewidth]{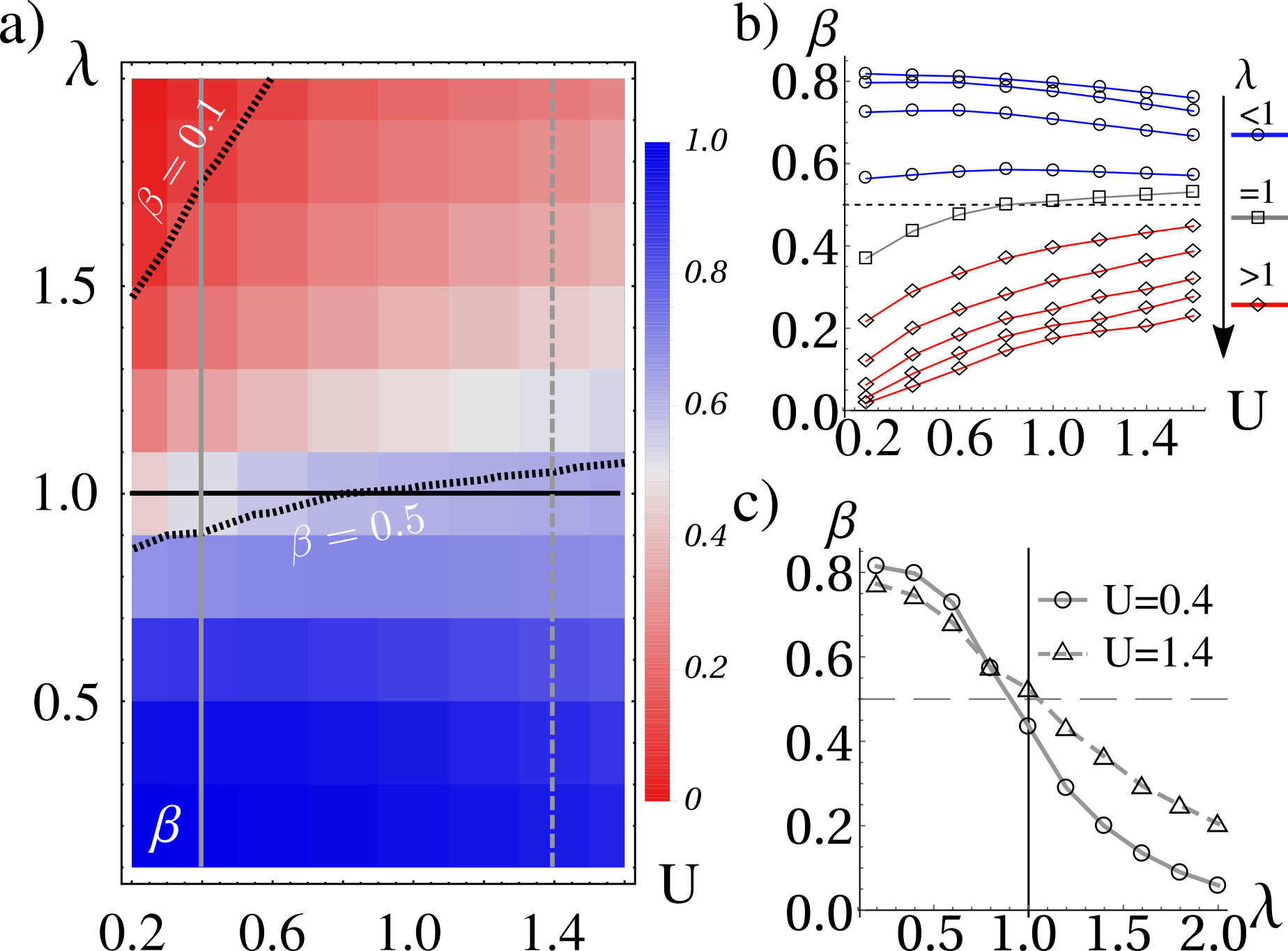}
	\caption{(Color online): a) Exponent of the power-law behaviour $\Delta N (t) \propto t^{-\beta}$ for the imbalance, in the AAM as a function of the potential strength $\lambda$ and interaction $U$. The black dashed curves identify $\beta=const$ levels, whereas the vertical grey lines point to the $U=const$ cuts shown in panel c). b) Exponent $\beta$ as a function of $U$ for different values of $\lambda$. The horizontal grey dashed line corresponds to $\beta=0.5$. c) Exponent $\beta$ for $U=0.4$ (solid) and $U=1.4$ (dashed) as a function of $\lambda$.}
	\label{fig:phdiag}
\end{figure}

Fig. ~\ref{fig:slowAA} {\bf a)} reports the imbalance $\Delta N(t)$ for the AAM and for $U=0.4$ and for different values of $\lambda$. The imbalance shows either a fast decay towards zero for $\lambda<1$ (with respect to the single particle time scale $J^{-1}$), or a slow decay, which, for higher $\lambda$, is also accompanied by persistent oscillations. This is a power-law decay, as we show in Fig. ~\ref{fig:slowAA} {\bf d)} (top). In order to assess this fact more quantitatively, we fitted~\cite{notefit} the imbalance with a power-law of the form $\Delta N(t)=  a t^{-\beta}$. The exponent $\beta$ for different $\lambda$ and $U$ is shown in Fig.\ref{fig:phdiag}. For $\lambda<1$ $\Delta N (t)\rightarrow 0$ in a super-diffusive way ($1/2 <\beta\lesssim 1$), and $\beta$ decreases with $U$, as expected for 1D systems at small interactions in the ergodic phase~\cite{Bloch2013}.
For $\lambda>1$, there are two appreciably different behaviours depending on the value of the interaction. A critical value $U_c(\lambda)$ exists, such that: for $U<U_c(\lambda)$ $\Delta N (t)\rightarrow \bar N\ne 0$,
$\beta\approx 0$, thus signalling long time localization; for $U\ge U_c(\lambda)$, $\Delta N (t)\rightarrow 0$ as a power-law, with an exponent smaller than that in the delocalized phase ($0 <\beta\lesssim 1/2$), showing a sub-diffusive behaviour.
In the latter parameter region, the time scale of the dynamics shows an anomalous dilation compared to the single particle one; but, still, this is very different from localization.

From Fig.\ref{fig:phdiag} {\bf c)}, one can also appreciate that our results are in good quantitative agreement with the ones extracted from the experiment in Ref.~\cite{Ulrich2017}.

\begin{figure}[t!]
	\vspace{0.5cm}
	\includegraphics[width=\linewidth]{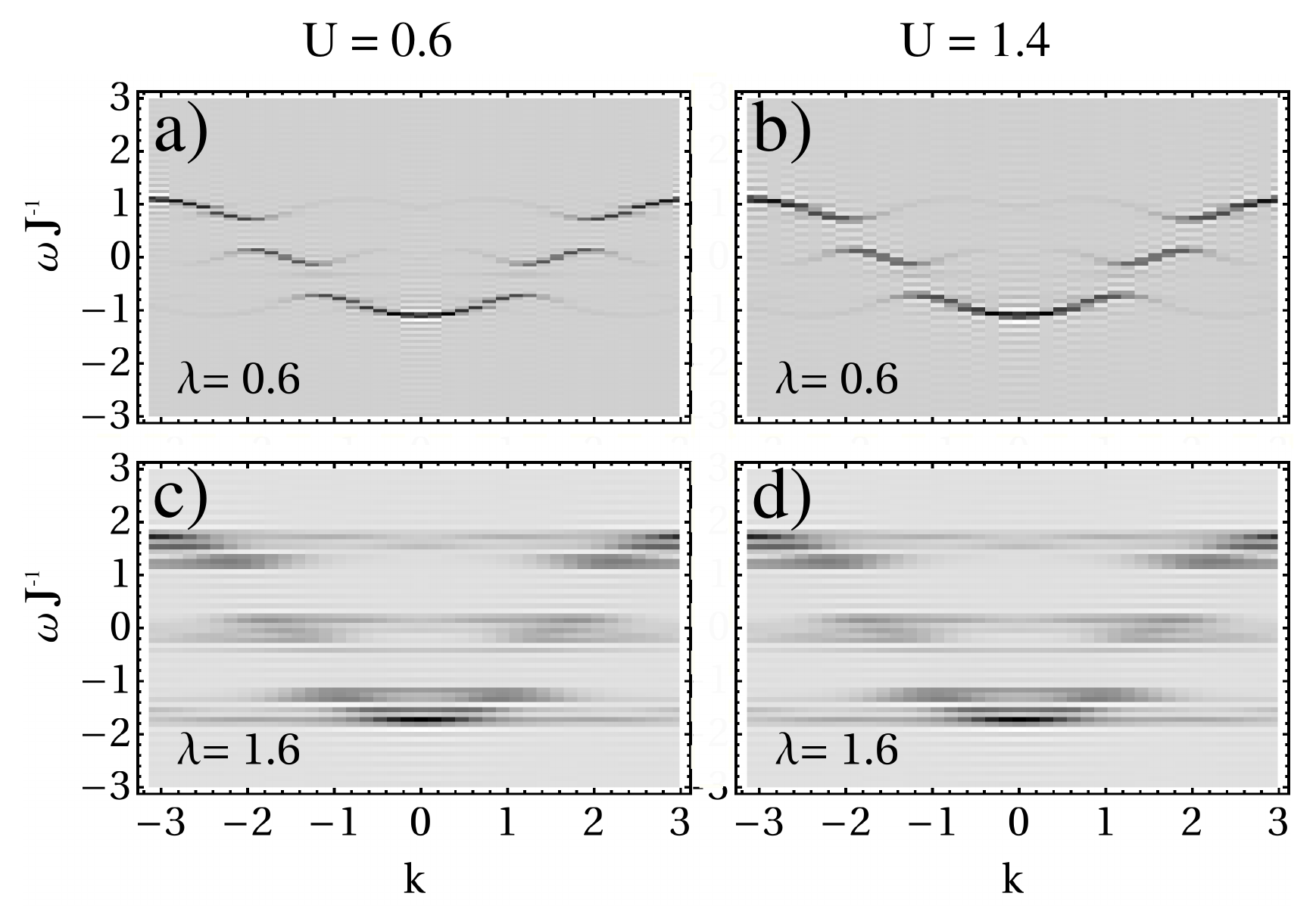}
	\caption{(Color online):  Spectral function $A(k,\omega)$ of the interacting AAM in the delocalized (panels (a) and
		(b)) and localized (panels (c) and (d)) phases.}
	\label{fig:spcfuncs}
\end{figure}

We want to highlight the fact that for such small system sizes having access to dynamical quantities is of paramount importance as normal equilibrium spectral properties might not reveal such features unless very large system sizes are accessible.
As an example in Fig.\ref{fig:spcfuncs} we plot the spectral function $A(k,\omega)$ for the AAM and for different points in the phase diagram in Fig.\ref{fig:phdiag}. Panel a) and b) refer to the case of a delocalized system. The different sub-bands induced by the modulation of the energy landscape are clearly visible. The two main gaps are clearly visible as well as other smaller ones in the top most sub-band and one in the lower one. The effect of the interaction is visible in the broadening of the peaks along the $\omega$ axis and in the closure of the smaller gaps.
This latter observation explains how the interaction induces the spectrum to become piece-wise continuous by closing the smallest gaps.
When the onsite potential $\lambda$ is higher (panels c) and d) in Fig.~\ref{fig:spcfuncs}) the spectral functions shows a broadening in momentum which corresponds to more localized states. Nevertheless the parameters of the system in panel d) correspond to an anomalous spreading ruling out the presence of localization.

\begin{figure}[h]
	\vspace{0.5cm}
	\includegraphics[width=\linewidth]{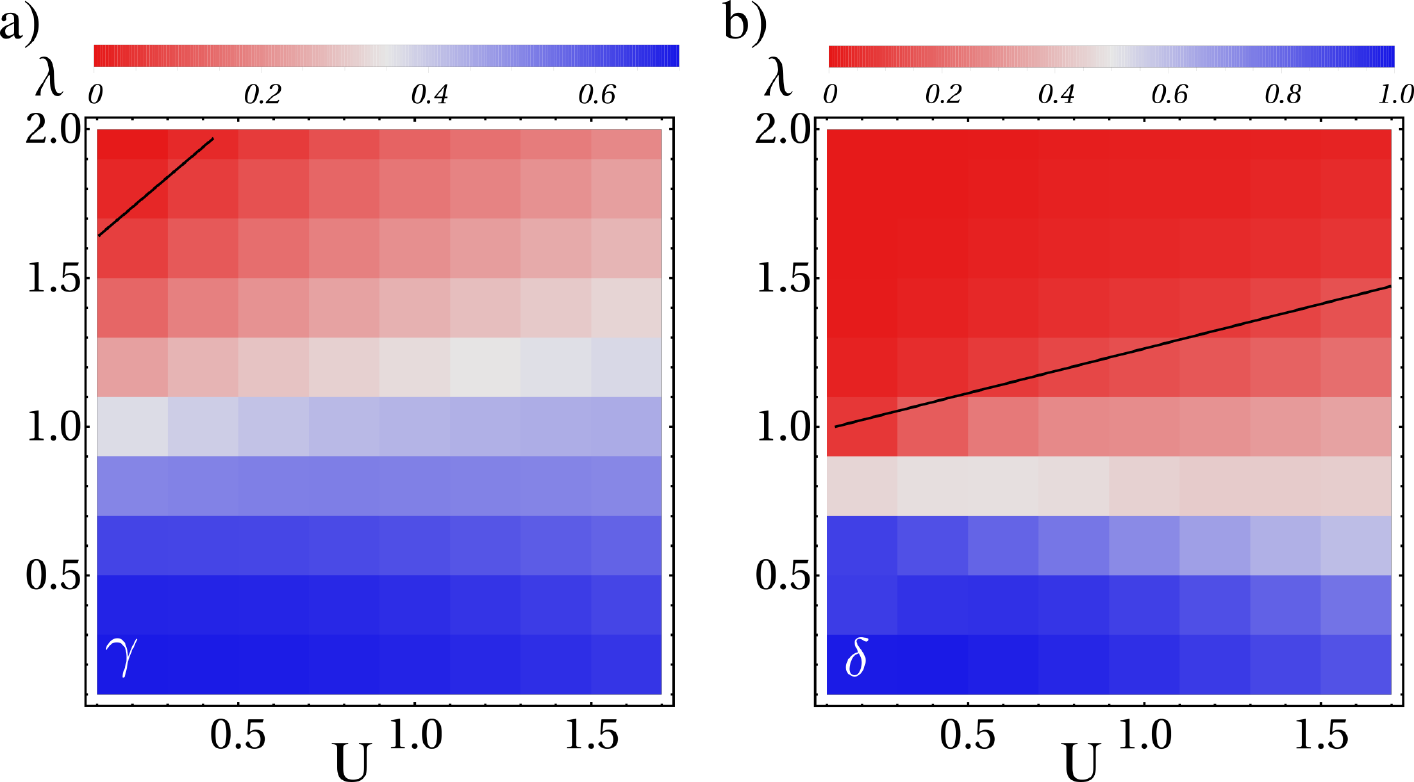}
	\caption{(Color online):   {\bf a)} Power law exponent for the decay at long times of the time-average $\langle \Delta N(t)\rangle_T\propto T^{-\gamma}$. The solid black line corresponds to $\gamma=0.1$. {\bf b)} Power law exponent for the decay at long times of the autocorrelation function $C(\tau)\propto \tau^{-\delta}$. The solid black line corresponds to $\delta=0.2$.
	}
	\label{fig:spectAA}
\end{figure}

\begin{figure*}[t!]
	\vspace{0.5cm}
	\includegraphics[width=\linewidth]{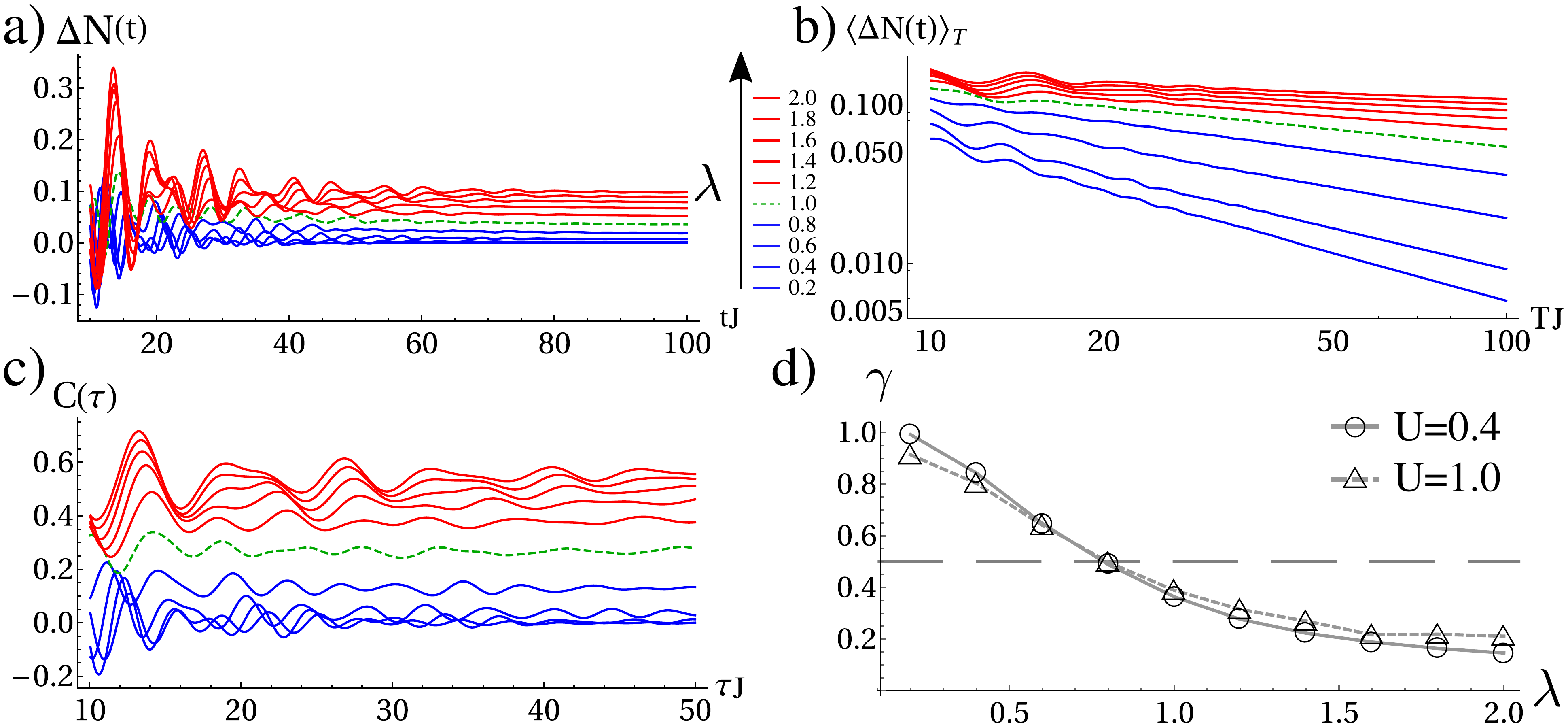}
	\caption{(Color online): {\bf a)} Particle imbalance between even and odd sites, $\Delta N(t)$, {\bf b)} Time-average $\langle \Delta N(t)\rangle_T$ (Log-Log), {\bf c)} Autocorrelation function for a system with $L=40$ sites in the interacting OFM for $U=0.4 $ for different values of $\lambda$.
		Panel {\bf d)} shows the exponent $\gamma$ of the power-law $\langle \Delta N(t)\rangle_T \propto T^{-\gamma}$ as a function of $\lambda$ for (solid grey with circles) $U=0.4$ and (dashed grey with triangles) $U=1.0$. The horizontal dashed line is at $\gamma=0.5$.}
	\label{fig:slowOF}
\end{figure*}
\section{SC spectrum in interacting systems} We conjecture that the slowing down of the dynamics observed above arises as a result of a non-trivial competition between the geometry of the underlying energy-landscape and the two-body interaction. To give a more solid ground to this conjecture, we shall show that the geometry-interaction interplay affects the nature of the SPES~\cite{notespec} and that there is a relation between the slowing-down and its SC nature.
Let us introduce the following quantities:
the time-averaged imbalance $\langle \Delta N(t)\rangle_T=T^{-1}\int\limits_{0}^{T} dt\;\; \Delta N(t)$ and the autocorrelation function $C(\tau) =\langle \Delta N^2(t) \rangle^{-1}  \langle \Delta N(t) \Delta N(t+\tau) \rangle$.

To investigate the nature of the SPES, we employ the results of Refs.\cite{Last1996,Pikovsky1995,Zhong1995,LoGullo2016}; specifically, we make use of the Ruelle-Amrein-Georgescu-Enss(RAGE) theorem and the Lebesgue-Riemann theorem, which imply that the conditions for the spectrum to have a SC component are $\lim_{T\rightarrow\infty}\langle \Delta N(t)\rangle_T=0 \wedge \lim_{\tau\rightarrow\infty} C(\tau)\neq 0$. The first condition excludes the presence of a PP part, whereas the second ensures that no AC part is present~\ref{app:spprt}.
For the data in Fig.\ref{fig:slowAA} {\bf a)}, these quantities are shown in Fig.\ref{fig:slowAA} {\bf b)}-{\bf c)}. In each of them, we can distinguish markedly different behaviors depending on the system parameters: a fast decay to zero, a slow decay towards zero and a decay towards a non-zero asymptotic value.
In Fig. \ref{fig:slowAA} {\bf d)} (bottom) we also show that $\langle \Delta N(t)\rangle_T$ has a power law behaviour, which also occurs for $C(\tau)$ (not shown). In Fig.~\ref{fig:spectAA}, we show the exponents of the power-law fits $\langle \Delta N(t)\rangle_T\propto T^{-\gamma}$ {\bf a)} and $C(\tau)\propto \tau^{-\delta}$ {\bf b)}.

We see that in the upper left region of Fig.~\ref{fig:spectAA} {\bf a)}, we have $\gamma<0.1$, which we conservatively assume as a threshold
for an almost non-decaying signal. According to the RAGE theorem, for the set of points below
this line, we can be sure that the SPES does not have a PP component.
Looking at the other exponent, $\delta$, in panel {\bf b} of Fig.~\ref{fig:spectAA}, we see that there is a region where $\delta<0.2$ (which is the exponent for $\lambda=1, U=0$, taken here, conservatively, as a threshold), for which $C(\tau)$ decays very slowly, and we expect the system not to have any AC component in its spectrum.
Merging these observations, we infer that in the region of parameters such that $\gamma>0.1$
and $\delta<0.2$ the spectrum of the system is purely SC.
It is important to highlight that the region where a SC component is present could be larger
than the one we are singling out, as we looked for regions where the spectrum is {\it purely} SC and tried to bound them accurately.

With the help of Fig.~\ref{fig:phdiag} {\bf b)}, we observe that there is
a good overlap between the region in which the anomalous slowing-down of $\Delta N(t)$
occurs and the region in which the system shows a SC SPES ($\{(\lambda,U)\;|\;\gamma>0.1 \text{ and } \delta<0.2\}$).

We now want to show that the observed time-scale-dilation is not a legacy of the transition at $U=0$, but it has a deeper origin. This fact emerges more clearly by looking at the behaviour of the interacting OFM, which gives rise to anomalous diffusion in the absence of interaction. In this model, the imbalance, reported in Fig.~\ref{fig:slowOF} for different $\lambda$ and $U=0.4$, shows a very slow decay. For small values of $\lambda$, $\Delta N(t)\rightarrow0$; whereas, for large $\lambda$, a power-law behaviour emerges similarly to the AAM. We can perform an analysis similar to the one conducted for the AAM on the SPES of the OFM. The time-averaged imbalance and the autocorrelation functions are shown in Fig.\ref{fig:slowOF} {\bf b)} and {\bf c)}. There, it can be clearly seen that $\langle \Delta N(t)\rangle_T$ decays to zero as a power law for all $\lambda$, whereas $C(\tau)$ reaches a constant value at large times. The behavior of the exponent $\gamma$ for two different interactions is shown in Fig.\ref{fig:slowOF} {\bf d)}, where one can see that increasing $\lambda$ results in a reduction of the decay exponent. These two observations show that the SPES of the interacting OFM is {\it purely} SC in nature. Actually, at small values of $\lambda$, we observe that the asymptotic value of $C(\tau)$ is zero. This does not rule out the presence of a SC component; but, instead, points towards the presence of the AC one. Nevertheless, for such small values of $\lambda$, the gaps induced by the underlying potential are very small and, therefore, any infinitely small interaction can cause their closure and the transition to a continuous of states.

The example of the OFM also shows that SC spectra are robust when many-body interactions are added, thus leaving hope of observing the unusual properties of quasicrystalline materials also in moderately interacting systems. This is in contrast with previous predictions~\cite{Kohomoto1992,LoGullo2015} based on effective non-interacting models, which allow us to conclude that many-body correlations are a key ingredient in the development of the discussed anomalous behavior.

\section{Conclusions}
In conclusion, we have shown that the anomalous slowing-down of the dynamics in the Aubry-Andr\`e model, observed in Ref.~\cite{Ulrich2017}, arises as a result of the singular continuous nature of the single particle energy spectrum. In the future, it will be interesting to investigate other models~\cite{Kohomoto1992} with a singular continuous spectrum in the absence of interactions and describe their fate when many-body interactions are introduced.

\begin{acknowledgments}
The authors acknowledge financial support from the Academy of Finland Centre of Excellence
program (Project no. 312058) and the Academy of Finland
(Project no. 287750).
NLG acknowledges financial support from the Turku Collegium for Science and Medicine (TCSM).
Numerical simulations were performed exploting the Finnish CSC facilties under the Project no. 2001004
("Quenches in weakly interacting ultracold atomic gases with non-trivial geometries").
\end{acknowledgments}

\appendix

\section{Single particle energy spectrum}
\label{app:spes}
We want to clarify what the meaning of ``single-particle energy spectrum'' used in the main
text in the case of an interacting many-body system. The quick definition is that it is the support of the density of states of the system.
To give a more explicit description, we will loosely follow the treatment given in Ref.\cite{stefleebook} (Chap. 6).
In the main text, we have chosen the local density as a figure of merit to analyse the spectral properties, which in terms of the Green's function is simply given by $n_j(t)=-i G_{jj}^<(t;t)$.
The latter can be written as:

\begin{equation}
G_{jj}^<(t;t)=i \left\langle e^{i \hat H t}\hat n_j e^{-i \hat H t} \right\rangle_{\hat \rho_0}
\end{equation}

We now introduce the identity operator:

\begin{equation}
\hat{ \mathds{1}}=\int_{\sigma} d\epsilon \;\ketbra{\Psi(\epsilon)}{\Psi(\epsilon)}
\end{equation}
where the integral is over the {\it whole} spectrum  $\sigma$ of the Hamiltonian $\hat H$, namely over the closure of the complement of the resolvent set, defined as $\rho=\{\lambda|(H-\lambda\mathds{1})^{-1} \text{  is a bounded operator}\}$ with respect to $\mathds{R}$.
According to the Lebesgue decomposition theorem, the spectrum is the union of three components $\sigma=\sigma_{ac}\cup\sigma_{sc}\cup\sigma_{pp}$ where ac, sc and pp stand for absolutely continuous, singular continuous, and pure point, respectively.
When $\epsilon$ belongs to the pp part of the spectrum, the integral notation is assumed to be replaced by a sum.
Inserting two identities into the expression for the lesser Green's function we obtain:
\begin{equation}
G_{jj}^<(t;t)=i \int_{\sigma} d\epsilon d\epsilon'\;\; e^{i  (\epsilon'-\epsilon) t} f_j(\epsilon,\epsilon')
\end{equation}
where $ f_j(\epsilon,\epsilon')=\braketM{\Psi(\epsilon)}{\hat \rho_0}{\Psi(\epsilon')}\braketM{\Psi(\epsilon')}{\hat n_j}{\Psi(\epsilon)}$.
The above expression can be recast into the form:

\begin{equation}
\label{eq:negl}
G_{jj}^<(t;t)=i \int_{-\infty}^{\infty}  e^{i\omega t} \;d\omega\;\mu_j(\omega)
\end{equation}
where we defined:
$\mu_j(\omega) =  \int_{\sigma} d\epsilon d\epsilon'\;\; \delta(\omega -(\epsilon'-\epsilon))\; f_j(\epsilon,\epsilon') $

In this form, the mean value of the the number of particle at site $j$
and time $t$ can be interpreted as the Fourier transform of a measure $\mu_j$, which has support on the spectrum of the total Hamiltonian $\hat H$.
Moreover, from the expression of $\mu_j$, we see that the measure is computed on the differences $\epsilon'-\epsilon$, namely it runs over all particle-hole-like excitations of the many-body system. In this respect, it can be seen as the single particle excitation spectrum.
To better understand this concept, let us look at a specific example.
Let us consider the case of a Fermi gas of N particles at zero temperature and at equilibrium, whose Hamiltonian is $\hat H_0$. If we now add a one-body perturbation, the total Hamiltonian reads $\hat H=\hat H_0+\delta \hat V$, where $\delta\hat V$ is a small perturbation.
Let us assume that, at time $t=0$, we suddenly switch this perturbation on (quantum quench).
We then expect that the explored spectrum will be that of all particle-hole excitations around the initial Fermi energy.

In the case of the initial state considered in the main text, we expect to explore most of the single-particle excitation spectrum as the initial state is a very highly excited one.

\section{Analysis of the spectral properties}
\label{app:spprt}
The link between the dynamics of the system and the nature of the single-particle energy spectrum can be highlighted by resorting to the theory of spectral analysis of operators.
It will be useful in the following to define the continuous component of a spectrum given by $\sigma_c=\sigma_{ac} \cup \sigma_{sc}$.

Let us introduce the RAGE (Ruelle-Amrein-Georgescu-Enss) theorem~\cite{Last1996,Reed1979}, which relates the time average of the mean of a compact operator to the presence of a continuous part. Given a compact operator $\hat A$ we define the time average of its expectation value at time $t$ as:

\begin{equation}
\langle\langle \hat A\rangle\rangle_T =\frac{1}{T}\int_0^T dt\; \langle \hat  A(t)\rangle_{\hat \rho_0}
\end{equation}

The RAGE theorem states that

\begin{equation}
\lim_{T\rightarrow \infty}\langle\langle A\rangle\rangle_T =0 \Leftrightarrow \sigma \subseteq\sigma_c
\end{equation}

The RAGE theorem gives a way to infer the presence of a pure-point component in the single-particle energy spectrum, which is guaranteed by the condition $\lim_{T\rightarrow \infty}\langle\langle A\rangle\rangle_T \neq 0$.

The number operator is a compact operator as it is a linear combination of projection operators; for the same reason, also the imbalance operator is a compact operator and, therefore, the RAGE theorem applies to the quantity $\langle \Delta N(t)\rangle_T$ considered in the main text.

The RAGE theorem alone still does not rule out the presence of an absolutely continuous part
whenever the time average goes to zero at long times.
In order to assess the presence (or absence) of the absolutely continuous part,
we look at the autocorrelation function:

\begin{equation}
C(\tau) =\frac{\left\langle \left\langle\hat A(t)\right\rangle_{\hat \rho_0}  \left\langle\hat A(t+\tau)\right\rangle_{\hat \rho_0} \right\rangle_t}{\left\langle \left\langle\hat A(t)\right\rangle_{\hat \rho_0}^2 \right\rangle_t}
\end{equation}

In the spectral analysis of signals, the autocorrelation functions provide a powerful method to asses the presence of correlations in time-series at different time lags, and, therefore, they can be  used to make statements on the nature of the spectrum without having direct access to the harmonic analysis of the signal itself.
Loosely speaking, if the spectrum has a pure point component, one expects sustained oscillations in the autocorrelation function showing order in the time.
The autocorrelation function will decay to zero, instead, 
if the signal is not correlated at long times, a feature to be expected in the presence of a continuous spectrum.
This physical intuition finds a more rigorous mathematical formulation, which will try to present briefly in the following.
It easy to see that in the case of the imbalance operator $\Delta \hat N\equiv\sum_{i}(-1)^i\hat n_i$ the autocorrelation function is given by :

\begin{equation}
C(\tau) =\int_{-\infty}^{\infty} e^{i \omega \tau} d\omega\;\;|f(\omega)|^2
\end{equation}

with $\mu(\omega)=\sum_{i}(-1)^i\mu_i(\omega)/N$ with $N=\int_{-\infty}^{\infty} d\omega\;\;|\sum_{i}(-1)^i \mu_i(\omega)|^2$.
Therefore, the autocorrelation function is nothing but the Fourier transform of a (positive) measure.
Comparing it with Eq.~\ref{eq:negl}, we see that this measure is the squared modulus of the sum of measures giving the occupation number at different sites.

Therefore, it turns out that the averaged autocorrelation function is nothing but the Fourier transform of the measure $d\omega |f(\omega)|^2$.
We can use its asymptotic behaviour to detect the presence of an absolutely continuous component of the spectrum.
Specifically, the Riemann-Lebesgue theorem tells us that $\lim_{\tau\rightarrow\infty}C(\tau)=0$ is a necessary condition for the spectrum to be {\it purely} absolutely continuous.
This means that $\lim_{\tau\rightarrow\infty}C(\tau)\neq 0$ implies that the spectrum is such that $\sigma \subseteq\sigma_s$ where $\sigma_s=\sigma_{pp}\cup\sigma_{sc}$ is the singular part
of the spectrum.

As a result, the conditions for the single particle excitation spectrum to be {\it purely} singular continuous can be written as:

\begin{align}
\label{eq:sc}
\lim_{T\rightarrow \infty}\langle\langle A\rangle\rangle_T &=0 \;\;\;\text{(no PP component is present)}\\
\lim_{\tau\rightarrow\infty}C(\tau)&\neq0\;\;\;\text{(no AC component is present)}
\end{align}

It is important to stress that, even in the case $\lim_{T\rightarrow \infty}\langle\langle A\rangle\rangle_T =0 \wedge \lim_{\tau\rightarrow\infty}C(\tau)=0$, a singular continuous component can still be present. This is due to the fact that from the Riemann-Lebesgue theorem
the condition $\lim_{\tau\rightarrow\infty}C(\tau)$ is necessary but not sufficient to guarantee
the presence of an AC component.
In this respect, the conditions ~\ref{eq:sc} to detect the presence of a singular continuous component are more strict than needed.


\begin{thebibliography}{99}

\bibitem[Shechtman D.(1984)]{Shechtman1984} D. Shechtman, I. Blech, D. Gratias, and J. W. Cahn.
\newblock Metallic Phase with Long-Range Orientational Order and No Translational Symmetry
\newblock {\em Phys. Rev. Lett.} {\bf 53}, 1951 (1984).

\bibitem[Tsai A.P.(1987)]{Tsai1987} A.P. Tsai, A. Inoue and T. Masumoto.
\newblock A Stable Quasicrystal in Al-Cu-Fe System.
\newblock {\em Jpn. J. Appl. Phys.} {\bf 26}, 1505 (1987).

\bibitem{dubois} J.-M. Dubois,
\newblock Useful Quasicrystals
\newblock (World Scientific, New Jersey, 2005).

\bibitem{note1} An eigenfunction is said to be critical if it is not delocalized nor exponentially localized; althought there exist different types of such eigenfunctions, most of them are characterized by a power law envelope and/or non-trivial (multi-)fractal properties.

\bibitem[Roche S.(1997)]{Roche1997}S.Roche, G. Trambly de Laissardi\`ere, and D. Mayou.
\newblock Electronic transport properties of quasicrystals.
\newblock {\em J. Matrh. Phys.}  {\bf 38}, 1794 (1997).

\bibitem{note2} According to the Lebesgue decomposition theorem a positive measure can be split into three (mutually orthogonal) components: absolutely continuous (AC), singular continuous (SC) and pure point (PP) according to the nature of their support.

\bibitem[Steinhardt P.J.(2013)]{Steinhardt2013}P.J. Steinhardt.
\newblock  Quasicrystals: a brief history of the impossible.
\newblock {\em Rendiconti Lincei}  {\bf 24}, 85 (2013).

\bibitem[Zhong J.X. (1995)]{Zhong1995}J.X. Zhong, and R. Mosseri.
\newblock Quantum dynamics n quasiperiodic systems.
\newblock {\em J. Phys.:Condens. Matter}  {\bf 7}, 8383 (1995).

\bibitem[Ketzmerick Y.(1992)]{Ketzmerick1992}R. Ketzmerick, G. Petschel, and T. Geisel.
\newblock Slow decay of temporal correlations in quantum systems with Cantor spectra.
\newblock {\em Phys. Rev. Lett.}  {\bf 69}, 695 (1992).

\bibitem[Lo Gullo N. (2017)]{LoGullo2017} N. Lo Gullo, C.V. Ambarish, Th. Busch, L. Dell'Anna, and C.M. Chandrashekar.
\newblock Dynamics and energy spectra of aperiodic discrete-time quantum walks.
\newblock {\em Phys. Rev. E} {\bf 96}, 012111 (2017).

\bibitem[Jitomirskaya S.Y.(1999)]{Jitomirskaya1999} S. Y. Jitomirskaya.
\newblock Metal-insulator transition for the almost Mathieu operator.
\newblock {\em Ann. of Math.}  {\bf 150}, 1159 (1999).

\bibitem[Jitomirskaya S.Y.(2009)]{Jitomirskaya2009} A. Avila and S. Y. Jitomirskaya.
\newblock The Ten Martini Problem.
\newblock {\em Ann. of Math.}  {\bf 170}, 303 (2009).

\bibitem{modugno} L. Tanzi, E. Lucioni, S. Chaudhuri, L. Gori, A. Kumar, C. D'Errico, M. Inguscio, and G. Modugno,
\newblock Transport of a Bose Gas in 1D Disordered Lattices at the Fluid-Insulator Transition.
\newblock {\em Phys. Rev. Lett.} {\bf 111}, 115301 (2013).

\bibitem[L\"uschen H. P.(2017)]{Ulrich2017}H.P. L\"uschen, P. Bordia, S. Scherg, F. Alet, E. Altman, U. Schneider, and I. Bloch.
\newblock Observation of slow dynamics near the many-body localization transition in one-dimensional quasiperiodic systems.
\newblock {\em Phys. Rev. Lett.} {\bf 119}, 260401 (2017).

\bibitem[Schreiber M.(2015)]{Ulrich2015}M. Schreiber, S.S. Hodgman, P. Bordia, H.P. Lschen, M.H. Fischer, R. Vosk, E. Altman, U. Schneider, I. Bloch
\newblock Observation of many-body localization of interacting fermions in a quasirandom optical lattice.
\newblock {\em Science} {\bf 349}, 842 (2015).

\bibitem[Roscilde T. (2008)]{Roscilde2008} T. Roscilde.
\newblock Bosons in one-dimensional incommensurate superlattices.
\newblock {\em Phys. Rev. A} {\bf 77}, 063605 (2008).

\bibitem[Deng X. (2008)]{Deng2008} X. Deng, R. Citro, A. Minguzzi, and E. Orignac.
\newblock Phase diagram and momentum distribution of an interacting Bose gas in a bichromatic lattice.
\newblock {\em Phys. Rev. A} {\bf 78}, 013625 (2008).

\bibitem[Roux G. (2008)]{Roux2008} G. Roux, T. Barthel, I. P. McCulloch, C. Kollath, U. Schollwock, and T. Giamarchi.
\newblock Quasiperiodic Bose-Hubbard model and localization in one-dimensional cold atomic gases.
\newblock {\em Phys. Rev. A} {\bf 78}, 023628 (2008).

\bibitem[Naldesi P.(2016)]{Naldesi2016} P. Naldesi, E. Ercolessi, T. Roscilde.
\newblock Detecting a many-body mobility edge with quantum quenches.
\newblock {\em SciPost Phys.} {\bf 1}, 010 (2016).

\bibitem[Settino J. (2017)]{Settino2017} J. Settino, N. Lo Gullo, A. Sindona, J. Goold, F. Plastina.
\newblock Signatures of the single-particle mobility edge in the ground-state properties of Tonks-Girardeau and noninteracting Fermi gases in a bichromatic potential.
\newblock {\em Phys. Rev. A} {\bf 95}, 033605 (2017).


\bibitem[Ancilotto F. (2018)]{Ancilotto2018} F. Ancilotto, D. Rossini, and S. Pilati.
\newblock Out-of-equilibrium dynamics of repulsive Fermi gases in quasiperiodic potentials: A density functional theory study.
\newblock {\em Phys. Rev. B} {\bf 97}, 155107 (2018).


\bibitem[Prelov\v sek P. (2016)]{Prelovsek2016} P. Prelov\v sek, O. S. Bari\v si\'c, and M. \=Znidari\v c.
\newblock Absence of full many-body localization in the disordered Hubbard chain.
\newblock {\em Phys. Rev. B}  {\bf 94}, 241104(R) (2016).

\bibitem[Mace N. (2018)]{Mace2018} N. Mac\'e, F. Alet, and N. Laflorencie.
\newblock Multifractal Scalings across the Many-Body Localization Transition.
\newblock Phys. Rev. Lett. {\bf 123}, 180601 (2019).

\bibitem{dassarma19} S. Xu, X. Li, Y.-T. Hsu, B. Swingle, S. Das Sarma,
\newblock Butterfly effect in interacting Aubry-Andr\'{e} model: Thermalization, slow scrambling, and many-body localization
Phys. Rev. Res. {\bf 1}, 032039(R) (2019).


\bibitem[Znidaric M. (2016)]{Znidaric2018} M. \=Znidari\v c, and M. Ljubotina.
\newblock Interaction instability of localization in quasiperiodic systems.
\newblock {\em PNAS}  {\bf 115}, 4595 (2018).

\bibitem[Kohomoto M. (1992)]{Kohomoto1992} H. Hiramoto, and M. Kohomoto.
\newblock Electronic Spectral and Wavefunction Properties of One-dimensional Quasiperiodic Systems: A Scaling Approach.
\newblock {\em Int. J. Mod. Phys. B}  {\bf 6}, 281 (1992).

\bibitem[Lo Gullo N. (2015)]{LoGullo2015} N. Lo Gullo, and L. Dell'Anna.
\newblock Spreading of correlations and Loschmidt echo after quantum quenches of a Bose gas in the Aubry-Andr\'e potential.
\newblock {\em Phys. Rev. A} {\bf 92}, 063619 (2015).

\bibitem{note3} We set the phase of the $\cos$ function to zero as it would not affect the results discussed in this work.
Indeed, its presence would have two main consequences: a reshuffling of the bulk eigenstates with respect to the eigenenergies and a change in the energy of the two (localized) boundary states, but neither of the two affects the SPES and its nature.

\bibitem[Talarico W.N. (2019)]{Talarico2019} W.N. Talarico, S. Maniscalco and N. Lo Gullo.
\newblock A scalable numerical approach to the solution of the Dyson equation for the non-equilibrium single-particle Green's function.
\newblock  	Phys. Status Solidi B {\bf 256}, 1800501 (2019).

\bibitem[Stan A. (2009)]{vanLeeuwen2009} A. Stan, N.E. Dahlen, and R. van Leeuwen.
\newblock Time propagation of the Kadanoff-Baym equations for inhomogeneous systems.
\newblock {\em J. Chem. Phys.} {\bf 130}, 224101 (2016).

\bibitem[Lynn R. A. (2016)]{vanLeeuwen2016} R.A. Lynn, and R. van Leeuwen.
\newblock Development of non-equilibrium Green's functions for use with full interaction in complex systems.
\newblock {\em J. Phys.: Conference Series} {\bf 696}, 012020 (2016).

\bibitem[Lo Gullo N. (2016)]{LoGullo2016} N. Lo Gullo, and L. Dell'Anna.
\newblock Self-consistent Keldysh approach to quenches in the weakly interacting Bose-Hubbard model.
\newblock {\em Phys. Rev. B} {\bf 94}, 184308 (2016).

\bibitem[Stefanucci G. (2013)]{stefleebook} G. Stefanucci and R. van Leeuwen.
\newblock Nonequilibrium Many-Body Theory of Quantum Systems
\newblock (Cambridge University Press, Cambridge, UK, 2013).

\bibitem[Lieb E. H (1972)]{Lieb1972} E.H. Lieb, and D.W. Robinson.
\newblock The finite group velocity of quantum spin systems.
\newblock {\em Commun. Math. Phys.} {\bf 28}, 251 (1972).

\bibitem{notelocexp} The residual expansion for $\lambda>1$ can be attributed to the tails of the exponentially localized eigenstates (due to the finite size). At $\lambda=1$ the exponent drops to $\alpha\approx1/2$, thus signaling deviation from both ballistic and localized behavior.

\bibitem[S\"uto A.(1989)]{Suto1989}A. Suto.
\newblock Singular continuous spectrum on a cantor set of zero Lebesgue measure for the Fibonacci Hamiltonian.
\newblock {\em J. Stat. Phys}  {\bf 56}, 525 (1989).

\bibitem[Lo Gullo N. (2016)]{LoGullo2016a} N. Lo Gullo, L. Vittadello, L. Dell'Anna, and M. Bazzan.
\newblock Equivalence classes of Fibonacci lattices and their similarity properties.
\newblock {\em Phys. Rev. A} {\bf 94}, 023846 (2016).

\bibitem[Lo Gullo N. (2017a)]{LoGullo2017a} N. Lo Gullo, L. Vittadello, L. Dell'Anna, M. Merano, N. Rossetto, and M. Bazzan.
\newblock A study of the brightest peaks in the diffraction pattern of Fibonacci gratings
\newblock {\em J. Opt.} {\bf 19}, 055613 (2017).

\bibitem[Queff\'elec M.(2010)]{Queffelec2010}M. Queff\'elec.
\newblock Substitution Dynamical Systems – Spectral Analysis: Second Edition,
\newblock {\em Lecture Notes in Mathematics}, 1294, Springer-Verlag Berlin Heidelberg 2010.

\bibitem[Luck J.M.(1989)]{Luck1989}J.M. Luck.
\newblock Cantor spectra and scaling of gap widths in deterministic aperiodic systems.
\newblock {\em Phys. Rev. B}  {\bf 39}, 5834 (1989).

\bibitem[Ronzheimer J. P.(2013)]{Bloch2013}J. P. Ronzheimer, M. Schreiber, S. Braun, S. S. Hodgman, S. Langer, I. P. McCulloch, F. Heidrich-Meisner, I. Bloch, and U. Schneider.
\newblock Expansion Dynamics of Interacting Bosons in Homogeneous Lattices in One and Two Dimensions.
\newblock {\em Phys. Rev. Lett.} {\bf 110}, 205301 (2013).

\bibitem{notefit} All fits have been performed by excluding the first few tunnelling times and specifically for $J t\ge 5$.
The same proedure has been applied in ref.~cite{Ulrich2017}. The reason is that the
initial, transient dynamics is ruled by single particle tunnelling.

\bibitem{notespec} Here by single particle spectrum we mean the single particle excitation spectrum. See~\ref{app:spes}.

\bibitem[Last Y.(1996)]{Last1996}Y. Last.
\newblock Quantum dynamics and decompositions of singular continuous spectra
\newblock {\em J. Funct. An.}  {\bf 142}, 406 (1996).


\bibitem[Pikovsky A.S.(1995)]{Pikovsky1995} A.S. Pikovsky, M.A. Zaks, U. Feudel, and J. Kurths.
\newblock Singular continuous spectra in dissipative dynamics.
\newblock {\em Phys. Rev. E}  {\bf 52}, 285 (1995).

\bibitem[Simon B. (1979)]{Reed1979} M. Reed, and B. Simon,
\newblock Methods of Modern Mathematical Physics, III. Scattering Theory.
\newblock London, San Diego: Academic Press (1979).

\bibitem[Amrein W.O. (1981)]{Amrein1981}W.O. Amrein.
\newblock Non-Relativistic Quantum Dynamics.
\newblock D. Reidel Publishing Company (1981).

\end{thebibliography}
\end{document}